\documentclass[prd,preprint,superscriptaddress,showkeys,showpacs,
 preprintnumbers,amsmath,amssymb,aps,floatfix]{revtex4-1}
\usepackage{graphicx}
\usepackage{epsfig}
\usepackage{amsfonts}
\usepackage{amsmath}
\usepackage{amssymb}
\usepackage{slashed}
\usepackage{rotating}
\usepackage{subfigure}
\usepackage{placeins}
\usepackage{type1cm}
\usepackage{dcolumn}
\usepackage{multirow}
\usepackage{bigstrut}
\usepackage{mathrsfs}
\usepackage[sort&compress]{natbib}
\usepackage{color}
\usepackage{setspace}

\newcommand{\bt}[1]{\textbf{#1}}

\begin{document}

\preprint{\today}

\title{Limits on Lorentz violation from charged-pion decay}

\author{J. P. Noordmans}
\affiliation{KVI, University of Groningen, Zernikelaan 25,
                   9747 AA Groningen, The Netherlands}

\author{K. K. Vos}
\affiliation{KVI, University of Groningen, Zernikelaan 25,
                   9747 AA Groningen, The Netherlands}
\affiliation{Centre for Theoretical Physics, University of Groningen, Nijenborgh 4,
                   9747 AG Groningen, The Netherlands}

\date{\today}
\vspace{3em}

\begin{abstract}
\noindent
Charged-pion decay offers many opportunities to study Lorentz violation. Using an effective field theory approach, we study Lorentz violation in the lepton, $W$-boson, and quark sectors and derive the differential pion-decay rate, including muon polarization. Using coordinate redefinitions we are able to relate the first-generation quark sector, in which no bounds were previously reported, to the lepton and $W$-boson sector. This facilitates a tractable calculation, enabling us to place bounds on the level of $10^{-4}$ on first-generation quark parameters. Our expression for the pion-decay rate can be used to constrain Lorentz violation in future experiments. 
\end{abstract}
%\keywords{}
\maketitle

\section{Motivation}
Many quantum-gravity theories predict scenarios in which Lorentz symmetry is (spontaneously) broken \cite{Lib09}. This breakdown of Lorentz symmetry is often studied in the context of the Standard-Model Extension (SME) \cite{Col98}. The SME is an effective field theory containing all possible Lorentz-violating terms that are singlets under the gauge group of the Standard Model (SM) of particle physics. These terms are built out of SM fields contracted with tensor coefficients that parametrize Lorentz violation. Many of these SME coefficients have been constrained with high precision~\cite{Kos11}, but weak decays still offer interesting possibilities to obtain new bounds, or to improve existing bounds \cite{Noo13a,Noo13b,Alt13a,Dia13}. 

In this paper we investigate Lorentz violation in the charged-pion decay $\pi\rightarrow \mu + \nu_\mu$. In our analysis we consider Lorentz violation in the lepton sector, the quark sector, and the $W$-boson propagator. We treat the quark-sector parameters in pion decay by using coordinate redefinitions, while we study the effects of the $W$-boson propagator in the context of a recently developed effective field theory framework~\cite{Noo13a,Noo13b}. This framework entails modifying the $W$-boson propagator with a general tensor $\chi^{\mu\nu}$, which can be related to a combination of gauge and Higgs coefficients in the SME. This approach has been applied to nuclear $\beta$~\cite{Noo13a,Noo13b,Wil13}, pion~\cite{Alt13}, kaon~\cite{Vos14}, and muon~\cite{Noo14} decays. We derive the Lorentz-violating differential decay rate, dependent on the direction of the outgoing muon and its polarization, which extends previous work~\cite{Alt11,Alt13}. Our results lead to many possibilities to constrain Lorentz-violating effects in future experiments. Using existing data we obtain bounds on first-generation quark coefficients in the SME.  

\section{Lepton parameters}
We start from the Lorentz-violating Lagrangian density for second-generation leptons, which contains the dimensionless and traceless coefficient $c^{\mu\nu}$,
\begin{equation}
\mathcal{L}^{\mathrm{LV}}_{\mathrm{leptons}} 
= c_{\mu\nu}\left[i\bar{\ell}\gamma^\mu \partial^\nu \ell + i\bar{\nu}\gamma^\mu \partial^\nu \nu
+ \bar{\ell}_L W^{\nu-}\gamma^\mu \nu_{L} + \bar{\nu}_L W^{\nu+}\gamma^\mu \ell_{L}\right]\ ,
\label{LVlagr}
\end{equation}
where $\ell$ is the charged-lepton field, $\nu$ is the neutrino field of the corresponding flavor, and $\psi_L = \tfrac{1}{2}(1-\gamma^5)\psi$.  We assume for simplicity that the coefficients for the left-handed  and right-handed fields are equal. Gauge invariance then dictates the equality of the neutrino and charged-lepton coefficients. To calculate the pion-decay rate from Eq.~\eqref{LVlagr} we follow the procedure developed in Ref.~\cite{Col01}. 

The Lagrangian density in Eq.~\eqref{LVlagr} contains additional, unconventional, time-derivative terms, which cause the Hamiltonian to be non-Hermitian in general. To remove these terms, we apply a field redefinition for both the neutrino and the charged lepton. This field redefinition is, to first order in Lorentz violation, given by~\cite{Blu97}
\begin{equation}
\psi = A\chi = \left(1-\tfrac{1}{2}c_{\mu 0}\gamma^0 \gamma^\mu\right)\chi\ ,
\label{fieldredef}
\end{equation}
where $\chi$ is the new physical field. Written in terms of $\chi$ the time-derivative term is conventional and the Hamiltonian is Hermitian.
In terms of the redefined fields, which we will write again as $\ell_L$ and $\nu_L$, the interaction term becomes
\begin{equation}
\mathcal{L} = W^{-}_\nu \bar{\ell}_L (g_{\mu\nu} + \mathcal{C}_{\mu\nu}) \gamma^\mu \nu_{L} = W^{-}_\nu \bar{\ell}_L \breve{\gamma}^\nu \nu_{L}\ ,
\label{redefintlagr}
\end{equation}
with
\begin{subequations}
\begin{eqnarray}
\breve{\gamma}^\mu & = & (g^{\mu\nu}+\mathcal{C}^{\nu\mu})\gamma_\nu\ , \label{muonvertex} \\
\mathcal{C}^{\mu\nu} & = & c^{\mu\nu} - c^{\mu 0}g^{0 \nu} + c^{\nu 0} g^{0 \mu} - c^{00}g^{\mu\nu}\ .
\end{eqnarray}
\end{subequations}
Hence, $\mathcal{C}^{\mu 0}=0$, which shows that the extra time-derivative terms have been removed by the field redefinition. From Eq.~\eqref{redefintlagr} we see that the vertex is now proportional to $\breve{\gamma}^\mu$, while it was proportional to $\gamma^\mu + c^{\nu\mu}\gamma_\nu$ before the redefinition.

From the Dirac equation, the dispersion relation and the spinor solutions can be obtained. When $c^{\mu\nu}$ is the only nonzero Lorentz-violating coefficient, the dispersion relation can be written as
$\tilde{p}^2 - \tilde{m}^2 = 0$,
with
$\tilde{p}^\mu = p^\mu + \mathcal{C}^{\mu\nu}p_\nu$ and $\tilde{m} = m(1-c^{00})$. 
The energy of both the particle and the antiparticle of either spin state is, to first order in Lorentz violation, given by
$E({\bf p}) = \bar{p}^0 - c_{\mu\nu}\bar{p}^\mu \bar{p}^\nu/\bar{p}^0$,
where we introduced the convenient notation $\bar{p} = (\bar{p}^0,{\bf p})$ with $\bar{p}^0 = \sqrt{{\bf p}^2+m^2}$. 
From the Dirac equation we determine that $u^s({\bf p})\bar{u}^s({\bf p})=(\slashed{\tilde{p}}+\tilde{m})(1+\gamma^5\slashed{\tilde{s}})/4\tilde{p}^0$
and $v^s({\bf p})\bar{v}^s({\bf p})=(\slashed{\tilde{p}}-\tilde{m})(1+\gamma^5\slashed{\tilde{s}})/4\tilde{p}^0$, with $\tilde{s}=\left(\frac{{\bf \tilde{p}}\cdot {\bf\hat{s}}}{\tilde{m}},{\bf \hat{s}} + \frac{({\bf \tilde{p}}\cdot {\bf\hat{s}}){\bf \tilde{p}}}{\tilde{m}(\tilde{m}+\tilde{p}^0)}\right)$, ${\bf \hat{s}}$ the muon spin in its restframe, and the spinors normalized to unity \cite{thesis}.
This results from explicit calculation or can be understood because $(\slashed{\tilde{p}} - \tilde{m})\chi=0$, which is just the normal Dirac equation with $p\rightarrow \tilde{p}$ and $m\rightarrow \tilde{m}$. %%%
We can now determine the squared matrix element for pion decay. After summing over neutrino spin and using momentum conservation, it is given by
\begin{equation}
\sum_{\nu\mathrm{\ spin}}|\mathcal{M}|^2  =
\frac{\tilde{m}^2G_F^2f_\pi^2}{\tilde{k}^0\tilde{p}^0}(\tilde{p}\pm\tilde{m}\tilde{s})\cdot\tilde{k}\ ,
\label{matrixelsq}
\end{equation}
where $p$ and $k$ are the muon and neutrino momentum, respectively, $G_F$ is the Fermi coupling
constant, $f_{\pi}\simeq 92$ MeV is the pion decay constant, and the upper (lower) sign applies for $\pi^-$ ($\pi^+$) decay. The matrix element is proportional to the muon mass. This can be understood by the usual spin-balance argument for pion decay, which shows that, in the pion restframe, the outgoing leptons should have the same helicity, while the weak interaction only couples to the chiral component of the charged-lepton field that is of the opposite handedness. Interestingly, the Lorentz-violating spinors are eigenvectors of the operator $\boldsymbol{\Sigma}\cdot{\bf \tilde{p}}$ instead of the usual helicity operator. Also, in the pion restframe, $\sum_{\nu\mathrm{\ spin}}|\mathcal{M}|^2 \propto (1\pm {\bf\hat{\tilde{p}}}\cdot {\bf\hat{s}})$, with ${\bf\hat{\tilde{p}}} = {\bf\tilde{p}}/|{\bf\tilde{p}}|$. This shows that the muons are polarized in the $\pm{\bf \tilde{p}}$-direction, instead of in the normal $\pm{\bf p}$-direction. This influences experiments that depend on pion decay for their polarized muons, such as $g-2$ \cite{gmin2} or TWIST \cite{twist}. %\com{right citation for TWIST? and PiBeta is not relevant?}. 
The first could detect the discussed effect, for example in the phase of the muon polarization, varying over the course of a sidereal day. Based on the current precision of the experiment a statistical precision of $10^ {-6}$ seems attainable.  

The differential decay rate is given by
\begin{equation}
d\Gamma = \frac{1}{2m_\pi}\frac{d^3p}{(2\pi)^3}\frac{d^3k}{(2\pi)^3}\sum_{\nu\mathrm{\ spin}}|\mathcal{M}|^2(2\pi)^4 \delta^4(q - p - k)\ ,
\label{gendifdecay}
\end{equation}
where $q$ is the pion momentum.
By using the dispersion relations and momentum conservation repeatedly, we find for the differential pion decay rate in the pion restframe
\begin{equation}
\frac{d\Gamma}{d\Omega} = \frac{G_F^2f_\pi^2}{8\pi^2}\widetilde{M}_-^2\left(\widetilde{M}_+ - \widetilde{M}_-\right)\left(1+3c^{00} + 3c^{ij}\hat{p}_i\hat{p}_j\right)\left(1\pm{\bf\hat{\tilde{p}}}\cdot{\bf \hat{s}}\right)\ ,
\label{cdecayprep}
\end{equation}
where $\hat{\tilde{p}}^i = \tilde{p}^i/|{\bf \tilde{p}}|= \hat{p}^i(1+c^{jk}\hat{p}_j\hat{p}_k)+c^{ij}\hat{p}_j$, $\widetilde{M}_+ = (m_\pi^2 + \tilde{m}^2)/2m_\pi$, $\widetilde{M}_- = (m_\pi^2 - \tilde{m}^2)/2m_\pi$, and Latin indices run over space indices only. We see that indeed the $\pi^-$ ($\pi^+$) decay rate vanishes if the muon spin is antiparallel (parallel) to ${\bf \hat{\tilde{p}}}$. We can write Eq.~\eqref{cdecayprep} more explicitly as
\begin{eqnarray}
\frac{d\Gamma}{d\Omega} &=& \frac{G_F^2 f_\pi^2}{8\pi^2} M_-^2(M_+ - M_-)
\Bigg[1 + c^{00} \frac{2M_+ - M_-}{M_-} + 3c^{ij}\hat{p}_i\hat{p}_j \notag \\
&& \qquad\qquad\qquad\qquad \pm ({\bf \hat{p}}\cdot{\bf \hat{s}})\left(1+c^{00}\frac{2M_+-M_-}{M_-}+4c^{ij}\hat{p}_i\hat{p}_j\right)\mp c^{ij}\hat{s}_i\hat{p}_j\Bigg]\ ,
\label{cdecay}
\end{eqnarray}
where $M_+$ and $M_-$ are equal to $\widetilde{M}_+$ and $\widetilde{M}_-$, with the replacement $\tilde{m}\rightarrow m$. There are no terms proportional to $c^{0j}\hat{p}_j$ or $c^{j0}\hat{p}_j$ in Eq.~\eqref{cdecay}. This implies that there
will be no difference in rate for muons going in opposite directions, when the polarization of the muons is not detected. Notice also that the decay rate, integrated over muon direction, does not depend on the muon spin, {\it i.e.} $\Gamma(\uparrow)-\Gamma(\downarrow)=0$. We expect this to be no longer the case when the coefficients for left-handed and right-handed fields are taken to be different. The energies of the two spin states are then no longer degenerate, and the form of the operators $u^s({\bf p})\bar{u}^s({\bf p})$ and $v^s({\bf p})\bar{v}^s({\bf p})$ is considerably more involved~\cite{thesis}.

\section{$W$-boson parameters}
Lorentz violation in pion decay can also result from the modified $W$-boson propagator
\begin{equation}
\left\langle W^{\mu+}W^{\nu-}\right\rangle = -i(g^{\mu\nu}+\chi^{\mu\nu})/M_W^2 \ ,
\label{wpropagator}
\end{equation}
where $\chi^{\mu\nu}$ parametrizes a broad class of Lorentz-violating effects in the SME~\cite{Noo13a}.
The difference with the conventional Lorentz-invariant calculation resides only
in the matrix element, which to first order in Lorentz violation reads
\begin{equation} 
\sum_{\nu\mathrm{\ spin}}\left|\mathcal{M}\right| = \frac{G_F^2 f_{\pi}^2}{2p^0 k^0} (g_{\mu\nu}g_{\rho\sigma}
+ \chi_{\mu\nu}g_{\rho\sigma} + g_{\mu\nu}\chi^*_{\rho\sigma})q^\mu q^\rho\,
\mathrm{Tr}\left[(\slashed{p}\mp m\slashed{s})\gamma^\nu\slashed{k}\gamma^\sigma(1-\gamma^5)\right]\ ,
\end{equation}
where $s$ is defined as $\tilde{s}$ with the replacements $\tilde{m}\rightarrow m$ and ${\bf \tilde{\hat{p}}}\rightarrow {\bf \hat{p}}$. Using Eq.~\eqref{gendifdecay} and performing the integrals over ${\bf k}$ and $|{\bf p}|$ results in
\begin{eqnarray}
\frac{d\Gamma}{d\Omega} &=& \frac{G_F^2 f_\pi^2}{8\pi^2}M_-^2
(M_+ - M_-) \Bigg[(1 \pm {\bf \hat{p}\cdot\hat{s}})(1+2\chi_r^{00} - 2\chi_r^{0j}\hat{p}_j\emph{})  \notag \\
&& \mp \frac{m_\pi}{m}\left[ 2\chi_r^{0j}\left(\hat{s}_j - \left({\bf \hat{p}\cdot\hat{s}}\right)\hat{p}_j\right)+ 2\chi_i^{0j}\left({\bf \hat{p}\times\hat{s}}\right)_j\right]\Bigg]\ ,
\label{decayrinchi}
\end{eqnarray}
where $\chi^{\mu\nu}_{r}$ and $\chi^{\mu\nu}_{i}$ denote the real and imaginary parts of $\chi^{\mu\nu}$, respectively. 

The differential decay rate with polarized muons in terms of $c^{\mu\nu}$
and $\chi^{\mu\nu}$ is now given in Eqs.~\eqref{cdecay} and \eqref{decayrinchi}, respectively. For $c^{\mu\nu}$
there are no terms proportional to $c^{0j}\hat{p}_j$, while for $\chi^{\mu\nu}$ there are no terms proportional
to $\chi^{ij}\hat{p}_i\hat{p}_j$. In the former case there will be a nonzero dipole asymmetry in the muon
direction, while one has to search for a higher-order multipole asymmetry in the latter case. Another difference between $c^{\mu\nu}$ and $\chi^{\mu\nu}$ is the enhancement factor $m_\pi/m$ for the spin-dependent terms in Eq.~\eqref{decayrinchi}, which is not present in Eq.~\eqref{cdecay}. For the dominant branching fraction $\pi\rightarrow \mu + \nu_\mu$ this is of order unity. However, if one would measure the electron spin in $\pi\rightarrow e + \nu_e$ decay, this gives a sizable enhancement. We point out that $\chi^{\mu\nu}$, in contrast with $c^{\mu\nu}$, produces a nonzero asymmetry in the spin of the muon:
\begin{equation}
\frac{\Gamma(\uparrow)-\Gamma(\downarrow)}{\Gamma(\uparrow)+\Gamma(\downarrow)} = \pm \frac{2}{3}\left(\frac{2m_\pi + m}{m}\right)\chi^{0z}_r\ ,
\end{equation}
where we chose the quantization axis in the $z$-direction. Finally, we notice that the decay rate in Eq.~\eqref{cdecay} has its maximum if ${\bf \hat{s}} = \pm {\bf \hat{\tilde{p}}}$. To first order in Lorentz violation Eq.~\eqref{decayrinchi} is proportional to $1\pm {\bf V}_\ell\cdot {\bf \hat{s}}$, with ${\bf V}_\ell$ given by $V^l_\ell = \hat{p}^l + 2m_\pi \left[\chi_r^{0l} + \hat{p}^l(\chi_r^{0j}\hat{p}_j)-\epsilon^{ljk}\hat{p}_j(\chi_i)_{0k}\right]/m$. Both $c^{\mu\nu}$ and $\chi^{\mu\nu}$ thus influence the polarization of the outgoing muons.

\section{Coordinate choices}
It is known \cite{Col02, Kos11b}, that some (combinations of) SME coefficients are physically unobservable. At the level of the Lagrangian, this can be shown by using field or coordinate redefinitions to bring the Lagrangian with the apparent Lorentz violation to a conventional Lorentz-symmetric form. Since the physics does not depend on a choice of coordinates or fields, the coefficients that can be removed are unobservable in experiments. In many cases interactions between different sectors of the SME prevent the full removal of the Lorentz-violating coefficients.

As an example of this, we look at a $c^{\mu\nu}$ parameter for a fermion field of a particular species, such as in Eq.~\eqref{LVlagr}. According to Refs.~\cite{Col02, Kos11b}, a Lagrangian with a nonzero $c^{\mu\nu}$ parameter is equivalent to a conventional Lagrangian in a skewed coordinate system. The $c^{\mu\nu}$ can be removed by a coordinate transformation $x^\mu \rightarrow x'^\mu = x^\mu + c^{\mu\nu}x_\nu$. However, this transformation introduces $-c^{\mu\nu}$ in the other fermion sectors, while for the gauge field sector $W^{\mu\nu}W_{\mu\nu} \rightarrow W^{\mu\nu}W^{\rho\sigma}\left(\eta_{\mu\rho}\eta_{\nu\sigma} + 2\eta_{\mu\rho}c_{\nu\sigma} + 2\eta_{\nu\sigma}c_{\mu\rho}\right)$. The latter has the same form as a partly nonzero $k^{\mu\nu\rho\sigma}$ parameter in the gauge field sector. 

If we only consider $c^{\mu\nu}$ coefficients for fermions and the relevant parts of the $k^{\mu\nu\rho\sigma}$ coefficients for gauge fields, we can always make one sector of the SME conventional by means of a coordinate transformation. Notice that this is not a general coordinate transformation, in the usual sense. This is because we do not transform the metric, but reinterpret the coordinates with respect to the metric. This means that we make a choice which sector of the Lagrangian defines the clocks and measuring rods, and is therefore the conventional sector. The choice as to which sector is conventional, depends on the experimental setup.

\section{Quark parameters}
We now turn to the quark sector of the SME. Although there are strict bounds
for effective parameters from meson oscillations and measurements on the neutron and the proton
\cite{Kos11}, the best bounds on actual quark parameters are in the top quark sector and they are at
the $10^{-1}$-$10^{-2}$ level \cite{Aba12}. Bounds on parameters for the other generations are lacking.
Using coordinate transformations, we calculate the effects of quark parameters in leptonic pion decay. 

The SM first-generation quark Lagrangian is given by
\begin{equation}
\mathcal{L}_{\mathrm{quark}} = \bar{u}(i\slashed{\partial}-m_u)u + \bar{d}(i\slashed{\partial}-m_d)d
 +\frac{g}{\sqrt{2}}V_{ud}\left[\bar{u}_L \slashed{W}^+ d_L + \bar{d}_L \slashed{W}^- u_L\right]\ ,
\end{equation}
where $g$ is the $SU(2)$ coupling constant and $V_{ud}$ is the relevant entry of the CKM matrix. The corresponding Lorentz-violating part of the SME Lagrangian is
\begin{equation}
\mathcal{L}_{\mathrm{quark}}^{\mathrm{LV}} = i c_{\mu\nu} \bar{u} \gamma^\mu \partial^\nu u
+ i c_{\mu\nu} \bar{d} \gamma^\mu \partial^\nu d + \frac{g}{\sqrt{2}}V_{ud}\left[c_{\mu\nu} \bar{u}_L \gamma^\mu W^{\nu+}d_L
+ c_{\mu\nu} \bar{d}_L\gamma^\mu W^{\nu-}u_L \right]\ ,
\label{LVquark}
\end{equation}
where we assume that Lorentz violation is equal for left-handed and right-handed quarks and that $c_{\mu\nu}$ is diagonal in flavor space. Gauge invariance then forces the parameters to be equal for up and down quarks. 

\begin{figure}[t]
	\centering
		\includegraphics[width=0.8\textwidth]{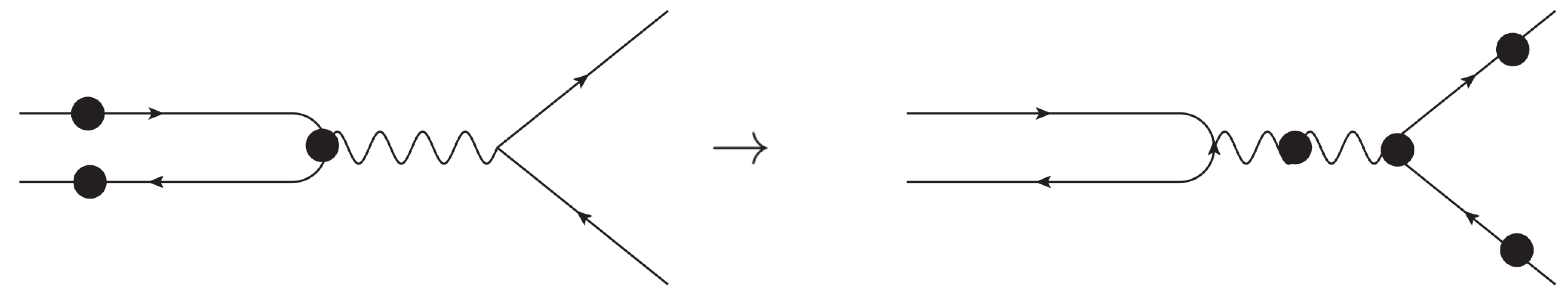}
	\caption{The effect of the coordinate transformations on pion decay. Blobs represent Lorentz violation.}
	\label{redef}
\end{figure}

As mentioned in the previous section, a coordinate transformation $x^\mu \rightarrow x'^{\mu} = x^{\mu} + c^{\mu\nu} x_{\nu}$ brings the quark Lagrangian to its conventional, Lorentz-symmetric, form. The coordinate transformation results in a low-energy $W$-boson propagator of the form in Eq.~\eqref{wpropagator} with $\chi^{\mu\nu} = 2c^{\mu\nu}$ and a $-c^{\mu\nu}$ coefficient for the second generation leptons. The effect of the coordinate transformations is visualized in the diagrams in Fig.~\ref{redef}. Notice that the transformation also changes the other sectors of the SME. It depends on the experimental conditions if this is relevant. In practice observables always depend on differences between Lorentz-violating parameters of the involved particles. Pion decay thus actually depends on differences between quark, lepton, and $W$-boson parameters. We will focus on the quark parameters the remainder of this paper. 

The calculation of the decay rate in terms of quark parameters can now be split in two parts, one dealing with Lorentz violation in the lepton kinetic terms and interaction vertex and one dealing with the Lorentz violation in the $W$-boson propagator. The former part of the
calculation exactly parallels the calculation above, with the substitution $c^{\mu\nu}\rightarrow -c^{\mu\nu}$. The latter part,
with a modified $W$-boson propagator, is treated by putting $\chi^{\mu\nu} = 2c^{\mu\nu}$. Since we treat Lorentz violation
to first order, we can simply combine the results in Eqs.~\eqref{cdecay} and \eqref{decayrinchi}, resulting in
\begin{eqnarray}
\frac{d\Gamma}{d\Omega} &=&  \frac{G_F^2 f_\pi^2}{8\pi^2}M_-^2(M_+ - M_-)
\Bigg[1 + c^{00}\frac{5M_- - 2M_+}{M_-} - 3c^{ij}\hat{p}_i\hat{p}_j - 4c^{0j} \hat{p}_j \notag \\
&& \pm({\bf \hat{p}}\cdot {\bf \hat{s}})\left[1+c^{00}\frac{5M_- - 2M_+}{M_-}-4c^{ij}\hat{p}_i\hat{p}_j + 4\left(\frac{m_\pi}{m}-1\right)c^{0j}\hat{p}_j\right]\mp \frac{4m_\pi}{m}c^{0j}\hat{s}_j \pm c^{ij}\hat{s}_i\hat{p}_j\Bigg]\notag\ . \\
\label{decayrquarkp}
\end{eqnarray}
To first order in Lorentz violating parameters, this decay rate is proportional to $1\pm {\bf V}_q\cdot{\bf \hat{s}}$, with $V_q^l = \hat{p}^l\left(1-c^{jk}\hat{p}_j\hat{p}_k\right) - c^{lk}\hat{p}_k + 4m_\pi\left[c^{0l}+\hat{p}^l(c^{0j}\hat{p}_j)\right]/m$, which summarizes the way the quark parameters will influence the polarization of the outgoing muons. The expression in Eq.~\eqref{decayrquarkp} offers many opportunities for future experiments to constrain the $c^{\mu\nu}$ quark coefficient, by observing the muon direction or spin in pion decay. 

%As discussed in the previous Section, experiments are always sensitive to combinations of Lorentz-violating parameters of different particle species, depending on the coordinate system one choses~\cite{Kos11b}. Depending on the experimental setup, Lorentz violating effects in the detection mechanism should then in principle also be taken into account, since they generally contain quarks. Other experiments might thus also be sensitive to Lorentz violation in the quark sector, but in general these effects are neglected in order to bound lepton parameters. For charged-pion decay we showed that the removal of the quark parameters is clean and tractable. Our calculation combined with some mild assumptions for the detection mechanism allows us to place bounds on the first-generation quark parameters. 

Integrating Eq.~\eqref{decayrquarkp} over muon directions and summing over spin gives the total decay rate
\begin{equation}
\Gamma/\Gamma_0 = 1+ (4M_- - 2M_+)c^{00}/M_-  \simeq (1-3.4 c^{00})\ .
\label{finalquark}
\end{equation}
Since this expression holds in the restframe of the pion, the sensitivity to Lorentz-violating effects in the decay rate is enhanced by a $\gamma_\pi^2$ dependence for pions in flight. Our result can be compared with the result in Refs.~\cite{Alt11,note} and the bounds in Refs.~\cite{Alt13a, Alt13}, derived from MINOS data~\cite{minos}. The translation of these bounds is complicated by possible Lorentz-violating effects in the detection system. As noted in Ref.~\cite{Alt13}, we expect from energy conservation that the processes in the detector are at least 4 times less sensitive to Lorentz violation. 
%For quark parameters, that might be involved due to atomic clock comparison, the sensitivity is expected to be reduced even further. 
Neglecting the effects of these processes and using the analysis in Ref.~\cite{Alt13a} together with Eq.~\eqref{finalquark}, we derive order-of-magnitude bounds on quark coefficients. These are listed in Table~\ref{table:qbounds}. The capital indices on {\it e.g.} $c^{TJ}$ denote time and space components in the standard Sun-centered inertial reference frame~\cite{Kos11}.
\begin{table}[t]
\setlength{\tabcolsep}{15pt} 
\begin{tabular}{|c|c|}
 \hline\hline Coefficient & Bound  \\ \hline
 $|c_{(TX)}|$, $|c_{(TY)}|$ & $4\times 10^{-5} $  \\
$|c_{XX}-c_{YY}|$, $| c_{(XY)}|$ & $ 9\times 10^{-5} $\\
  $| c_{(XZ)}|$, $| c_{(YZ)}|$& $ 7\times 10^{-5} $ \\ \hline\hline
\end{tabular}
\caption{3$\,\sigma$ bounds on $u$, $d$ quark parameters from the analysis in Ref.~\cite{Alt13a}, which uses MINOS data \cite{minos}; $c_{(TJ)}\equiv c_{TJ}+c_{JT}$.}
\label{table:qbounds}
\end{table}

From Eq.~\eqref{finalquark} bounds on the isotropic components of the quark tensor $c^{\mu\nu}$ can also be found, by using the ratio of the decay rates for $\pi \rightarrow e + \nu_e$ and $\pi \rightarrow \mu + \nu_\mu$. For the $\pi \rightarrow e + \nu_e$ rate, we need to remember that the electron sector is also modified by the coordinate transformations. This results in a decay rate as in Eq.~\eqref{finalquark} with the electron mass replacing the muon mass. The ratio then becomes
\begin{equation}
R_{\pi} \equiv \frac{\Gamma(\pi^{-}\rightarrow e^{-}\bar{\nu}_e)}
{\Gamma(\pi^{-}\rightarrow \mu^{-}\bar{\nu}_\mu)} = (1+5.4 c^{TT})R_{\pi}^{\textrm{SM}} \ , % as for electron (21) = (1+2c00)
\end{equation}
where $R_{\pi}^{\textrm{SM}} = 1.2352(1) \times 10^{-4}$ is the theoretical SM value \cite{Cir07} and $R_{\pi}=1.230(4)\times 10^{-4}$
is the experimental value \cite{Ber12}. By attributing the deviation from the SM value to $c^{TT}$ we find
\begin{equation}
c^{TT} = -8(6) \times 10^{-4} \ .
\label{cttbound}
\end{equation}
The same method can also set bounds at the $10^{-3}$-$10^{-4}$ level on the isotropic parts of $\chi^{\mu\nu}$, but this precision is not competitive with the precision in Ref.~\cite{Noo13b}. For the muon $c^{TT}$ coefficient the method gives a value of $c^{TT}=6(4)\times 10^ {-4}$, which, to our knowledge, is the first bound on this parameter.

\section{Discussion}
In this paper we obtained bounds on first-generation quark parameters, summarized in Table~\ref{table:qbounds} and Eq.~\eqref{cttbound}. We calculated the Lorentz-violating differential pion decay rate with polarized muons. The results for Lorentz violation in the lepton sector, the $W$-boson propagator, and the quark sector are given in Eqs.~\eqref{cdecay}, \eqref{decayrinchi}, and \eqref{decayrquarkp}, respectively. These offer many experimental opportunities to improve bounds on Lorentz violation. We also noted some qualitative differences between the influence of the different parameters on the pion decay rate. These pertain to asymmetries in muon and muon-spin directions, enhancements of spin effects for $\pi \rightarrow e + \nu_e$, and unusual polarization directions of the outgoing muons.

Weak decays have been used in the past to obtain direct bounds on Lorentz violation in the lepton and gauge sector.  
Bounds on $\chi^{\mu\nu}$, derived from forbidden $\beta$ decay, are at the $10^{-6}$ level on $\chi^{TT}$ and $\chi^{TJ}$,
and are on the order of $10^{-8}$ for $\chi^{JK}$  \cite{Noo13b}.
The analysis of pion decay using MINOS data \cite{minos} limits $c^{TJ}$ and $c^{JK}$
to the level of $10^{-5}$~\cite{Alt13a}.

A dedicated experiment for pion decay that measures the muon direction can provide improved
bounds, in particular on $\chi^{TJ}$ and $c^{\mu\nu}$ for both the lepton and the quark sector. Such an experiment should preferably benefit from the $\gamma_\pi^2$ dependence, as this increases the sensitivity of the measurement to Lorentz-violating effects and reduces uncertainties arising from possible Lorentz violation in the detection mechanism. We can estimate the reachable precision of such an experiment. Pion beams with an intensity of $10^{10}$/s are available at modern facilities,
suggesting a reachable precision on muon flux asymmetries of $10^{-4}\sqrt{\textrm{s}}$ or better. A statistical
precision of the order of $10^{-6}$ on various lifetime asymmetries thus seems attainable, enabling a dedicated experiment to put competitive or new bounds on the
Lorentz-violating parameters in lepton, quark, and gauge sectors of the SME.

\section{Acknowledgments}
We thank B. Altschul, A. Kosteleck\'y, G. Onderwater, and R. Timmermans for helpful discussions. This research was supported by the Stichting voor Fundamenteel
Onderzoek der Materie (FOM) under programmes 104 and 114 and project 08PR2636.


\begin{thebibliography}{99}
\bibitem{note} The total decay rate, that follows from Eq.~\eqref{cdecay}, differs from the one in Ref.~\cite{Alt11} by a factor of $(1-c_{00})^2$. This can be traced to the field redefinition in Eq.~\eqref{fieldredef} and the corresponding normalization of the fields. Our approach agrees with Ref.~\cite{Col01}.
\end{thebibliography}
\end{document}